\documentclass[a4paper,11pt]{article}
\pdfoutput=1
\usepackage{jheppub}

\usepackage[utf8]{inputenc}

\allowdisplaybreaks
\usepackage{amsmath}
\usepackage{amssymb}
\usepackage{amsmath}
\usepackage{breqn}
\usepackage{braket}
\usepackage{xcolor}
\usepackage{graphicx}
\usepackage{slashed}
\usepackage{appendix}
\usepackage{placeins}

\allowdisplaybreaks

\newcommand{\ud}{\mathrm{d}}

\def\as{a_s}

\preprint{{MSUHEP-23-024}, ZU-TH 43/23}

\title{Complete $N_f^2$ contributions to four-loop pure-singlet splitting functions}

\author[1]{Thomas Gehrmann,}
\emailAdd{thomas.gehrmann@uzh.ch}
\author[2, 3]{Andreas von Manteuffel,}
\emailAdd{manteuffel@ur.de}
\author[1]{Vasily Sotnikov,}
\emailAdd{vasily.sotnikov@physik.uzh.ch}
\author[1, 3]{Tong-Zhi Yang}
\emailAdd{toyang@physik.uzh.ch}

\affiliation[1]{Physik-Institut, Universität Zürich, Winterthurerstrasse 190, 8057 Zürich, Switzerland}
\affiliation[2]{Institut für Theoretische Physik, Universität Regensburg, 93040 Regensburg, Germany}
\affiliation[3]{Department of Physics and Astronomy, Michigan State University, East Lansing, MI 48824, USA}

\abstract{
The scale evolution of parton distributions 
is determined by universal splitting functions. 
As a milestone towards the computation of 
these functions to four-loop order in QCD, we compute all contributions to the pure-singlet quark-quark splitting functions that involve two closed fermion loops. The 
splitting functions are extracted from the 
pole terms of off-shell operator matrix elements, and the workflow for their calculation is outlined.
We reproduce known results for the non-singlet 
four-loop splitting functions and validate our new 
pure-singlet 
results against fixed Mellin moments. 
}


\begin{document}

\maketitle

\section{Introduction}
\label{sec:introduction}

The parton distribution functions are 
universal ingredients to the theory prediction of 
any collider observable that involves hadrons 
in the initial state. Their scale evolution~\cite{Altarelli:1977zs,Gribov:1972ri,Dokshitzer:1977sg} is 
governed by splitting functions, which can be 
determined order-by-order in perturbative QCD. 
Complete analytic expressions for 
the splitting functions are available up to three loops in QCD~\cite{Altarelli:1977zs,Gross:1974cs,Floratos:1978ny,Gonzalez-Arroyo:1979qht,Curci:1980uw,Furmanski:1980cm,Moch:2004pa,Vogt:2004mw}.

To enable consistent
predictions for collider observables,  a
matching level of precision is required 
among the
hard subprocess cross sections and the 
parton evolution: NLO QCD predictions involve 
parton distributions evolved according to the two-loop 
splitting functions and NNLO QCD implies three-loop 
evolution. Following 
pioneering results on the Higgs and Drell-Yan cross sections~\cite{Anastasiou:2015vya,Mistlberger:2018etf,Duhr:2020seh}, an increasing number of N3LO calculations 
for benchmark collider processes 
is now being accomplished (see e.g.~\cite{Heinrich:2020ybq} for a recent review). These highlight the 
urgent need for four-loop parton evolution. 
Partial results at four loops 
were obtained for parts of the 
 non-singlet splitting functions~\cite{Moch:2017uml}
and for a finite number of Mellin moments of the 
singlet splitting functions~\cite{Moch:2021qrk,Falcioni:2023luc,Falcioni:2023vqq}. 
This information on the four-loop splitting functions 
could already be used to 
approximate N\textsuperscript{3}LO-accurate parton distributions~\cite{McGowan:2022nag,Hekhorn:2023gul}. 
The computation of the full set of the four-loop 
splitting functions remains an outstanding task, 
required to enable fully consistent hadron 
collider predictions at N3LO accuracy. 

In this paper, we employ the framework for 
computing the splitting functions from 
the divergences of massless off-shell operator matrix elements (OMEs). This technique is based on the operator product expansion (OPE) and has been applied successfully in splitting function 
calculations at lower loop orders~\cite{Gross:1974cs,Floratos:1978ny,Gonzalez-Arroyo:1979qht,Hamberg:1991qt,Blumlein:2021enk,Blumlein:2021ryt,Gehrmann:2023ksf}.  
Compared to the extraction of splitting functions 
from physical subprocess coefficient functions~\cite{Moch:2004pa,Vogt:2004mw}, the 
OPE-based approach is particularly attractive, since it typically leads to simpler types of Feynman amplitudes and Feynman integrals. However, the off-shell nature of the operator matrix elements gives rise to a complicated mixing between the physical operators 
and unphysical gauge-variant operators under renormalization~\cite{Dixon:1974ss,Kluberg-Stern:1974iel,Collins:1994ee,Falcioni:2022fdm}. 

In this paper, we describe the calculation of
all contributions with two closed fermion loops 
to the pure-singlet splitting functions at four loops, 
using the OPE technique. 
To determine the renormalization counterterms resulting from these gauge-variant operators we follow the novel procedure proposed by some of us in~\cite{Gehrmann:2023ksf}, which is 
summarized in Section~\ref{sec:renormalizationTwist2}. 
The computation of OMEs to four-loop order is described
in Section~\ref{sec:methods}. This workflow is then applied in Section~\ref{sec:results} to obtain the 
results for the complete four-loop $N_f^2$
contributions to the pure-singlet splitting functions and 
to confirm previous results for the $N_f^2$ contributions to the non-singlet 
splitting functions. We conclude with an outlook in Section~\ref{sec:outlook}.

\section{Renormalization of the twist-two operators}
\label{sec:renormalizationTwist2}
To study the collinear behavior of QCD at leading power, we consider the twist-two operators from the operator product expansion. With regards to the flavor group, the twist-two operators are decomposed into non-singlet and singlet parts. The non-singlet operators of spin $n$ are given by
\begin{equation}
\label{eq:nonsingletOP}
O^{\mu_1 \cdots \mu_n}_{q,k} = \frac{i^{n-1}}{2}  \mathcal{S} \bigg[ \bar{\psi}_{i_1} \gamma^{\mu_1} D^{\mu_2}_{i_1 i_2} D^{\mu_3}_{i_2 i_3}\cdots D^{\mu_n}_{i_{n-1} i_n} \,\frac{\lambda_k}{2} \psi_{i_n} - \text{trace terms} \bigg],\, k = 3, \cdots N_f^2-1 \,, 
\end{equation}  
while the two singlet quark and gluon operators are 
\begin{align}
\label{eq:singletOP}
O^{\mu_1 \cdots \mu_n}_q &= \frac{i^{n-1}}{2}  \mathcal{S} \bigg[ \bar{\psi}_{i_1} \gamma^{\mu_1} D^{\mu_2}_{i_1 i_2} D^{\mu_3}_{i_2 i_3}\cdots D^{\mu_n}_{i_{n-1} i_n} \psi_{i_n} - \text{trace terms} \bigg] \,, \nonumber  \\
O^{\mu_1 \cdots \mu_n}_g &=-\frac{i^{n-2}}{2} \mathcal{S} \bigg[ G^{\mu_1}_{a_1, \mu} D^{\mu_2}_{a_1 a_2} \cdots D^{\mu_{n-1}}_{a_{n-2} a_{n-1}} G^{\mu_n \mu}_{a_{n-1} a_n}  - \text{trace terms} \bigg] \,,  
\end{align}
where $\mathcal{S}$ denotes symmetrization of Lorentz indices $\mu_1 \cdots \mu_n$ and $\lambda_k/2$ are diagonal generators of the flavor group $\text{SU}(N_f)$. In the above equations, $\psi$ and $G_{\mu \nu}^a$ represent the quark field and gluon field strength tensor, respectively, and $D^\mu = \partial_\mu  -i g_s  \textbf{T}^a  A^a_\mu $ is the covariant derivative in either the fundamental or the adjoint representation of a general gauge group. In this paper, we consider only the operators associated with zero-momentum transfer.

In practice, it is convenient to extract the information of interest by contracting the above operators with a fully symmetric external source
$J_{\mu_1 \cdots \mu_n} = \Delta_{\mu_1} \Delta_{\mu_2} \cdots \Delta_{\mu_n}$, where $\Delta$ is light-like with $\Delta^2 = 0$. In this way, we define the following spin-$n$ operators,
\begin{align}
O_{q,k} &= O^{\mu_1 \cdots \mu_n}_{q,k}  J_{\mu_1 \cdots \mu_n} \,, \nonumber \\ 
O_{q} &= O^{\mu_1 \cdots \mu_n}_{q}  J_{\mu_1 \cdots \mu_n} \,, \nonumber \\ 
O_{g} &= O^{\mu_1 \cdots \mu_n}_{g}  J_{\mu_1 \cdots \mu_n} \,.
\end{align}

As usual, the non-singlet operator is renormalized multiplicatively, i.e.,
\begin{equation}
\label{eq:nsRenormalization}
O^{\text{R}}_{q,k}  = Z_{\text{ns}} O^{\text{B}}_{q,k} \,,
\end{equation} 
where here and below we introduce the superscript R and B to denote the renormalized and bare operators, respectively. 

However, for singlet operators, in addition to mixing among themselves under renormalization, they also mix with other unphysical operators, the so-called gauge-variant (GV) operators. This mixing was first pointed out by Gross and Wilczek in the first extraction of the one-loop singlet anomalous dimensions~\cite{Gross:1974cs}. Subsequently, the renormalization of twist-two operators and the theory of the renormalization of a general gauge-invariant operator have been widely studied in the literature, by the seminal works of Dixon and Taylor~\cite{Dixon:1974ss}, Kluberg-Stern and Zuber~\cite{Kluberg-Stern:1974nmx,Kluberg-Stern:1975ebk}, Joglekar and Lee~\cite{Joglekar:1975nu}, Collins and Scalise~\cite{Collins:1994ee}. These works allowed explicit derivations of the GV operators at order $g_s$ and enabled the extraction of the correct anomalous dimension to two loops~\cite{Hamberg:1991qt}, thereby resolving earlier inconsistencies between different groups~\cite{Floratos:1978ny,Gonzalez-Arroyo:1979qht}. However, it is not clear how to generalize those works to enable the construction of the GV operators beyond order $g_s$. Recently, starting from a generalized gauge symmetry and promoting it to a generalized BRST symmetry, Falcioni and Herzog~\cite{Falcioni:2022fdm} were able to construct the GV operators for fixed $n$ to higher orders.

A general framework has been formulated to derive the renormalization counterterms resulting from the GV operators with all-$n$ dependence~\cite{Gehrmann:2023ksf}. For completeness, we describe this framework briefly below, the details can be found in~\cite{Gehrmann:2023ksf}. The framework generalizes the naive renormalization
\begin{align}
\label{eq:mixingOqOg}
\left( \begin{array}{c} 
O_q \\
O_g
\end{array} \right)^{\text{R,naive}} =  \left( \begin{array}{cc} 
Z_{qq} & Z_{qg} \\
Z_{gq} & Z_{gg} 
\end{array} \right)   \left( \begin{array}{c} 
O_q \\
O_g 
\end{array} \right)^{\text{B}}\,
\end{align}
to 
\begin{align}
\label{eq:mixingOqOgOA}
\left( \begin{array}{c} 
O_q \\
O_g \\
O_{ABC}\\
\end{array} \right)^{\text{R}} =  \left( \begin{array}{ccc} 
Z_{qq} & Z_{qg} & Z_{qA} \\
Z_{gq} & Z_{gg} & Z_{gA}  \\ 
0 & 0 & Z_{AA}  \\
\end{array} \right)   \left( \begin{array}{c} 
O_q \\
O_g \\
O_{ABC}\\
\end{array} \right)^{\text{B}} + 
\left( 
\begin{array}{c}
\left[Z O\right]_{q}^{\textrm{GV}}\\
\left[Z O\right]_{g}^{\textrm{GV}}\\
\left[Z O\right]_{A}^{\textrm{GV}}
\end{array}
\right)^{\text{B}}\,,
\end{align}
where we introduce GV operator $O_{ABC} = O_A + O_B + O_C$ with $O_A,\, O_B,\,O_C$ denoting the gluon, quark, and ghost GV operators respectively.  As discussed in~\cite{Gehrmann:2023ksf}, the GV operator $O_{ABC}$ alone is insufficient for renormalizing the physical operators $O_q$ and $O_g$. Additional terms are necessary, denoted in the above equation as $\left[Z O\right]_{q}^{\textrm{GV}}$, $\left[Z O\right]_{g}^{\textrm{GV}}$, and $\left[Z O\right]_{A}^{\textrm{GV}}$, where $Z$ and $O$ are written together. This notation is used because it becomes impractical to disentangle the renormalization constants from their associated operators while retaining the complete dependence on all powers of $n$. Thus, the renormalization constants here should be distinguished from those appearing in the first term on the right-hand side of the above equation. Notice that these additional terms are GV counterterms for the purpose of canceling the non-physical contributions only. Thus it is not necessary for the above equation to exhibit the pattern of multiplicative renormalization. In addition to the physical renormalization constants $Z_{ij}$ with $i,j=q,g$, we also introduce non-physical renormalization constants $Z_{qA},Z_{gA}, Z_{AA}$ associated with GV operators $O_{ABC}$. These non-physical renormalization constants and GV counterterms only start to contribute from a certain order in $a_s = g_s^2/(4 \pi)^2$, specifically: 
\begin{alignat}{2}
    Z_{qA} &= \mathcal{O}(a_s^2),\, &\qquad \left[Z O\right]_{q}^{\textrm{GV}} &= \mathcal{O}(a_s^3) \,, \nonumber 
    \\
    Z_{gA} & =   \mathcal{O}(a_s),\, &\left[Z O\right]_{g}^{\textrm{GV}} &= \mathcal{O}(a_s^2)\,,
\end{alignat}
Our strategy~\cite{Gehrmann:2023ksf} is to explicitly extract the counterterm Feynman rules associated to the GV operators instead of determining the GV operators themselves. The strategy relies on considering multi-leg, off-shell operator matrix elements (OMEs), which are defined as the Green's functions or matrix elements with an operator insertion. For example, in the two-point case we have, 
\begin{align}
\label{eq:OMEsDe}
A^{}_{ij} = \braket{j(p)|O_i|j(p)}^{} \text{ with } p^2<0\,,
\end{align} 
where $O_i$ represents a twist-two operator and $j$ with momentum $p$ denotes a quark, gluon or ghost external state. The established framework is valid to all orders in $g_s$ and we have worked out in~\cite{Gehrmann:2023ksf} the Feynman rules for $O_{ABC}$ to order $g_s^2$ as well as the counterterm Feynman rules for $\left[ZO\right]_{g}^{\textrm{GV}}$ to order $a_s^2 \,g_s$, where $a_s$ and $g_s$ stemming from $Z$ and $O$ respectively. These counterterm Feynman rules are enough to extract physical renormalization constants as well as $Z_{qA},\, Z_{gA}$ to three-loop order. As we will see below, the counterterm 
Feynman rules derived in~\cite{Gehrmann:2023ksf} are sufficient to determine $Z_{qq}$ to four-loop order. 

To extract $Z_{qq}$, we only need to consider the renormalization of $O_q$, 
\begin{eqnarray}
O^{\text{R}}_q &=& Z_{qq} O^{\text{B}}_q + Z_{qg} O^{\text{B}}_g + Z_{qA} O^{\text{B}}_{ABC}+  \left[Z O\right]_{q}^{\textrm{GV}}\,. 
 \label{eq:RenorGVSimQ}
 \end{eqnarray} 
 Inserting the above equations into two-quark external states, we obtain
 \begin{align}
\label{eq:2qomeForOqRenor}
     \braket{q|O^{\text{R}}_q|q} &= Z_{q} \bigg[ Z_{qq} \braket{q|O^{\text{B}}_q|q} \nonumber
     \\
     &+ Z_{qg} \braket{q|O^{\text{B}}_g|q} + Z_{qA} \braket{q|O^{\text{B}}_{ABC}|q}+  \braket{q|\left[Z O\right]_{q}^{\textrm{GV}}|q} \bigg]\bigg|_{a_s^{\text{B}} \to Z_{a_s} a_s,\, \xi^{\text{B}} \to Z_g \xi } \,,
 \end{align}
 where we introduced the quark and gluon wave function renormalization constants $Z_q$, $Z_g$, and the strong coupling renormalization constant $Z_{a_s}$. Explicit expressions for them are collected in appendix \ref{app:qcdRenor}. Further, $\xi$ is the gauge parameter, where $\xi =1$ corresponds to Feynman gauge. 

 For the determination of  $Z_{qq}$ at four loops, 
 we need to compute the OME $\braket{q|O^{\text{B}}_q|q}$ to four-loop order, which will be described in detail in Section~\ref{sec:methods} below. In addition we use the known results for OMEs $\braket{q|O^{\text{B}}_g|q}$ and $\braket{q|O^{\text{B}}_{ABC}|q}$ up to three-loop and two-loop orders respectively~\cite{Gehrmann:2023ksf}. Lastly, the OME $\braket{q|\left[Z O\right]_{q}^{\textrm{GV}}|q}$ needs to be evaluated to four-loop order. It was shown in the appendix of~\cite{Gehrmann:2023ksf}, that the counterterm Feynman rules for the $qq$, $gg$ and $qqg$ vertices resulting from $\left[Z O\right]_{q}^{\textrm{GV}}$ are zero, which leads to the following conclusion: 
 \begin{equation}
     \braket{q|\left[Z O\right]_{q}^{\textrm{GV}}|q} = \mathcal{O}(a_s^5)\,.
 \end{equation}
 
 The renormalization constants above satisfy the renormalization group equations
\begin{equation}
\label{eq:renormalizaitonEq}
 \frac{d Z }{d \ln \mu} = -2  \gamma_{ } \cdot Z  \,.
\end{equation}
In the non-singlet case, the anomalous dimension can be extracted from $Z_{\text{ns}}$ by solving~\eqref{eq:renormalizaitonEq} with the help of the $d$-dimensional QCD $\beta$ function
\begin{align}
 \beta(\as,\,\epsilon) = \frac{d \as }{ d \ln \mu } = -2 \epsilon \, \as - 2 \as \sum_{i=0}^{\infty} \as^{i+1} \beta_i \,,
\end{align}
where $\epsilon = (4-d)/2$.
Explicitly, we have,
\begin{align}
\label{eq:ZfactorIntermsofGammaNS}
Z_{\text{ns}} = & 1 + a_s \frac{\gamma^{(0)}_{\text{ns}}}{\epsilon}+a_s^2 \Bigg( \frac{\gamma_{\text{ns}}^{(1)}}{2 \epsilon} + \frac{1}{2 \epsilon^2} \bigg[ -\beta_0 \gamma_{\text{ns}}^{(0)} +  \big( \gamma_{\text{ns}}^{(0)}\big)^2  \bigg] \Bigg) \nonumber 
\\
& + a_s^3  \Bigg( \frac{1}{3 \epsilon} \gamma_{\text{ns}}^{(2)} + \frac{1}{6 \epsilon^2} \bigg[ -2 \beta_1 \gamma_{\text{ns}}^{(0)} - 2 \beta_0 \gamma_{\text{ns}}^{(1)}+ 3  \gamma^{(0)}_{\text{ns}} \gamma^{(1)}_{\text{ns}}    \bigg] \nonumber 
\\
& \quad + \frac{1}{6\epsilon^3} \bigg[ 2 \beta_0^2 \gamma_{\text{ns}}^{(0)}  - 3 \beta_0  \big( \gamma_{\text{ns}}^{(0)} \big)^2  + \big( \gamma_{\text{ns}}^{(0)} \big)^3   \bigg] \Bigg) \nonumber 
\\
&+ \frac{a_s^4}{24} \Bigg( \frac{1}{\epsilon^4} \bigg[-6 \beta _0^3 \gamma_{\text{ns}}^{(0)}-6 \beta _0 (\gamma _{\text{ns}}^{(0)})^3+11 \beta _0^2 (\gamma _{\text{ns}}^{(0)})^2+(\gamma
   _{\text{ns}}^{(0)})^4 \bigg] \nonumber 
   \\
   & \quad + \frac{1}{\epsilon ^3} \bigg[  {6 \beta _0^2 \gamma_{\text{ns}}^{(1)}-14 \beta _0 \gamma_{\text{ns}}^{(0)} \gamma_{\text{ns}}^{(1)}+12 \beta _0 \beta _1 \gamma_{\text{ns}}^{(0)}+6 \gamma_{\text{ns}}^{(1)} (\gamma
   _{\text{ns}}^{(0)})^2-8 \beta _1 (\gamma _{\text{ns}}^{(0)})^2}{} \bigg] \nonumber 
   \\
   & \quad +\frac{1}{\epsilon ^2}\bigg[{-6 \beta _0 \gamma_{\text{ns}}^{(2)}-6 \beta _1 \gamma_{\text{ns}}^{(1)}-6 \beta _2 \gamma_{\text{ns}}^{(0)}+8
   \gamma_{\text{ns}}^{(0)} \gamma_{\text{ns}}^{(2)}+3 (\gamma _{\text{ns}}^{(1)})^2}{}\bigg]+\frac{6 \gamma_{\text{ns}}^{(3)}}{\epsilon } \Bigg) \nonumber 
   \\
   &+ 
 \mathcal{O}(a_s^5)\,.
\end{align}
 The non-singlet anomalous dimension $\gamma_{\text{ns}}$ can be decomposed into the following form by separating the even and odd moments,
\begin{align}
    \gamma_{\text{ns}} = \frac{1+(-1)^n}{2} \gamma_{\text{ns}}^+ + \frac{1-(-1)^n}{2} \left(  \gamma_{\text{ns}}^- + \gamma^{\text{s}}_{\text{ns}} \right), 
\end{align}
where the detailed definitions of $\gamma_{\text{ns}}^{\pm}$ and $\gamma_{\text{ns}}^{\text{s}}$ can be found, for example, in~\cite{Moch:2004pa}. Here and in the rest of this paper, we always expand the anomalous dimension according to 
\begin{equation}
\gamma  = \sum_{l=1}^\infty a_s^l \gamma^{(l-1)}\,, 
\end{equation}
while for the renormalization constants we follow a different convention,
\begin{equation}
\label{eq:zexpandConvention}
Z  = \sum_{l=0}^\infty a_s^l Z^{(l)}\,. 
\end{equation}
In the singlet case, our goal is the determination of the physical anomalous dimensions $\gamma_{ij}$.
They can be read off from the physical renormalization constants $Z_{ij}$,   
\begin{align}
\label{eq:ZfactorIntermsofGamma}
Z_{ij} = &+\delta_{ij} + \as \frac{\gamma^{(0)}_{ij}}{\epsilon}+\as^2 \Bigg( \frac{\gamma_{ij}^{(1)}}{2 \epsilon} + \frac{1}{2 \epsilon^2} \bigg[ -\beta_0 \gamma_{ij}^{(0)} + \sum_{k=q,\,g} \gamma^{(0)}_{ik} \gamma^{(0)}_{kj}  \bigg] \Bigg) \nonumber 
\\
& + \as^3  \Bigg( \frac{1}{3 \epsilon} \gamma_{ij}^{(2)} + \frac{1}{6 \epsilon^2} \bigg[ -2 \beta_1 \gamma_{ij}^{(0)} - 2 \beta_0 \gamma_{ij}^{(1)} +2 \sum_{k=q,\,g} \gamma^{(1)}_{ik} \gamma^{(0)}_{kj}  +\sum_{k} \gamma^{(0)}_{ik} \gamma^{(1)}_{kj}    \bigg] \nonumber 
\\
& \quad + \frac{1}{6\epsilon^3} \bigg[ 2 \beta_0^2 \gamma_{ij}^{(0)}  - 3 \beta_0 \sum_{k=q,\,g} \gamma_{ik}^{(0)} \gamma_{kj}^{(0)} + \sum_{k=q,\,g} \sum_{l=q,\,g} \gamma^{(0)}_{ik} \gamma^{(0)}_{kl} \gamma^{(0)}_{lj}   \bigg] \Bigg) \nonumber 
\\
&+ \frac{a_s^4}{24} \Bigg( \frac{1}{\epsilon ^4} \bigg[11 \beta _0^2 \sum_{k=q,\,g}\gamma ^{(0)}_{ik}\gamma ^{(0)}_{kj}-6 \beta _0 \sum_{k=q,\,g} \sum_{l=q,\,g}\gamma
   ^{(0)}_{ik}\gamma ^{(0)}_{kl}\gamma ^{(0)}_{lj} \nonumber 
   \\
   & \quad +\sum_{k=q,\,g} \sum_{l=q,\,g} \sum_{m=q,\,g}\gamma
   ^{(0)}_{ik}\gamma ^{(0)}_{kl}\gamma ^{(0)}_{lm}\gamma ^{(0)}_{mj}-6 \beta _0^3 \gamma
   ^{(0)}_{ij} \bigg] +\frac{1}{{\epsilon ^3}} \bigg[-5 \beta _0 \sum_{k=q,\,g} \gamma ^{(0)}_{ik}\gamma ^{(1)}_{kj}\nonumber 
   \\
   & \quad -9
   \beta _0 \sum_{k=q,\,g}\gamma ^{(1)}_{ik}\gamma ^{(0)}_{kj}-8 \beta _1 \sum_{k=q,\,g} \gamma
   ^{(0)}_{ik}\gamma ^{(0)}_{kj}+\sum_{k=q,\,g} \sum_{l=q,\,g} \gamma ^{(0)}_{ik}\gamma
   ^{(0)}_{kl}\gamma ^{(1)}_{lj} \nonumber 
   \\
   &\quad +2 \sum_{k=q,\,g} \sum_{l=q,\,g}  \gamma ^{(0)}_{ik}\gamma
   ^{(1)}_{kl}\gamma ^{(0)}_{lj} +3 \sum_{k=q,\,g} \sum_{l=q,\,g}\gamma ^{(1)}_{ik}\gamma
   ^{(0)}_{kl}\gamma ^{(0)}_{lj}+6 \beta _0^2 \gamma ^{(1)}_{ij}+12 \beta _1 \beta _0 \gamma
   ^{(0)}_{ij}\bigg] \nonumber 
   \\
   & \quad +  \frac{1}{\epsilon ^2} \bigg[2 \sum_{k=q,\,g}\gamma ^{(0)}_{ik}\gamma ^{(2)}_{kj}  +3
   \sum_{k=q,\,g} \gamma ^{(1)}_{ik}\gamma ^{(1)}_{kj}+6 \sum_{k}\gamma ^{(2)}_{ik}\gamma
   ^{(0)}_{kj}-6 \beta _2 \gamma ^{(0)}_{ij}\nonumber 
   \\
   & \quad -6 \beta _1 \gamma ^{(1)}_{ij}-6 \beta _0 \gamma
   ^{(2)}_{ij}{} \bigg]  +\frac{6 \gamma ^{(3)}_{ij}}{\epsilon }    \Bigg)  + \mathcal{O}(a_s^5)\,,
\end{align}
where $i,j=q,g$. Notice that if we consider only $q$ instead of $g$ for the dummy indices $k,l,m$ in the above summations, the above equation has the same form as the equation~\eqref{eq:ZfactorIntermsofGammaNS}. 
Furthermore, equation~\eqref{eq:ZfactorIntermsofGamma} remains unaltered even when GV operators (counterterms) are introduced. This is due to the fact that the renormalization of GV operators (counterterms) does not involve mixing with the physical operators, as demonstrated in~\cite{Joglekar:1975nu}. In other words, the mixing matrix in~\eqref{eq:mixingOqOgOA} has a block-triangular form. In this paper, we are mainly interested in the $N_f^2$ contributions to both, the non-singlet anomalous dimension $\gamma^{(3)}_{\text{ns}}$ and the pure-singlet anomalous dimension $\gamma^{(3)}_{\text{ps}}$ defined by
\begin{equation}
    \gamma^{(3)}_{\text{ps}} = \gamma^{(3)}_{qq} - \gamma^{+, (3)}_{\text{ns}}\,,
\end{equation}
where $N_f$ is the number of massless quark flavors.

\section{Computational method}
\label{sec:methods}
\begin{figure}
\begin{center}
\begin{minipage}{7.5cm}
\includegraphics[scale=0.95]{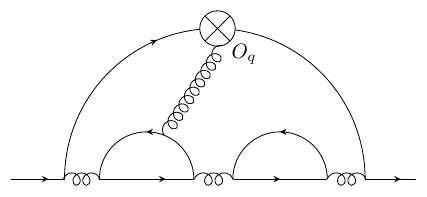} 
\end{minipage}
\begin{minipage}{7.5cm}
\includegraphics[scale=0.95]{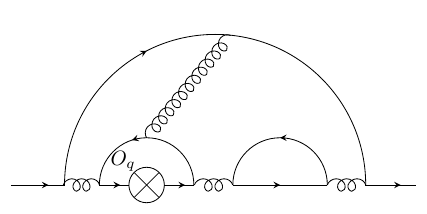} 
\end{minipage}
\caption{Representative Feynman diagrams for $N_f^2$ contributions to the OME $\braket{q|O^{\text{B}}_{q}|q}$ at four loops. The first diagram contributes to the non-singlet anomalous dimension, while the second diagram contributes to the pure-singlet anomalous dimension in the quark channel.}
\label{fig:diagrams}
\end{center}
\end{figure}

As demonstrated above, for the purpose of extracting $Z_{qq}$ at the four-loop order, the last missing contribution is the OME $\braket{q|O^{\text{B}}_q|q}$ at four loops. In this section, we focus on the computation of the $N_f^2$ part for this OME. The corresponding Feynman diagrams were generated by \texttt{QGRAF}~\cite{Nogueira:1991ex}, see Fig.~\ref{fig:diagrams} for some sample Feynman diagrams. The required Feynman rules in $n$-space 
involve non-standard terms like $(\Delta \cdot p)^{n-1}$ and are not convenient for 
the application of integration-by-parts (IBP) reductions~\cite{Chetyrkin:1981qh,Tkachov:1981wb,Laporta:2000dsw}. To overcome this problem, a method first proposed in~\cite{Ablinger:2012qm,Ablinger:2014nga} is adopted. The method sums the non-standard terms into linear propagators depending on a tracing parameter $t$. For example,
\begin{align}
\label{eq:sumXmethod}
( \Delta \cdot p)^{n-1} \to \sum_{n=1}^\infty ( \Delta \cdot p)^{n-1} t^n = \frac{t}{1- t \Delta \cdot p}\,.
\end{align}
In the following, we always work in  parameter-$t$ space, which allows us to use standard IBP algorithms. To generate the unreduced amplitude in parameter-$t$ space, Mathematica is used to substitute the Feynman rules into the Feynman diagrams generated by \texttt{QGRAF}~\cite{Nogueira:1991ex} and \texttt{FORM}~\cite{Vermaseren:2000nd} is used to deal with the Dirac and color algebra. To reduce the size of the raw amplitude, we first classify the diagrams into different integral families with an in-house code invoking \texttt{Reduze 2}~\cite{vonManteuffel:2012np} and \texttt{FeynCalc}~\cite{Shtabovenko:2016sxi,Shtabovenko:2021hjx}. During the family classification, partial fractions among the Feynman propagators are needed, especially for the linear propagators, and \texttt{Apart}~\cite{Feng:2012iq} is used for this task.
At later stages of the calculation, we also use \texttt{MultivariateApart}~\cite{Heller:2021qkz} to decompose rational functions of kinematic invariants and parameters into multivariate partial fractions (see also \cite{Pak:2011xt,Bendle:2021ueg,Gerlach:2022qnc} for alternative decompositions).

To reduce the amplitude and to calculate the master integrals in the differential equation approach~\cite{Gehrmann:1999as}, we perform IBP reductions of the four-loop integrals using finite field sampling and rational reconstruction~\cite{vonManteuffel:2014ixa,Peraro:2016wsq,Peraro:2019svx}. An optimized input system of equations is prepared by employing the method of~\cite{Agarwal:2020dye} to control the generation of squared propagators.
We note that the unreduced amplitude contains not only irreducible numerators but also higher powers of the propagators, in particular for the linear ones.
For the computation of the reductions, we group integrals into so-called sectors, which label different sets of denominators.
For each sector, we generate IBP identities by constructing suitable differential operators and subsequently applying them to so-called seed integrals which have positive powers of the denominators of the sector.
We eliminate redundant equations from the system for each sector, 
where for performance reasons we ignore relations which involve only integrals in sub-sectors, that is, with fewer different denominators.
It is well known that ignoring subsector information in this way can lead to incomplete reductions, since one misses specific relations, which are sometimes referred to as ``hidden'' or ``anomalous''.
In the present case, we recover such relations from differential equations and dimensional analysis, as will be explained below.
The private code \texttt{Finred} is used to perform the filtering as well as the final reduction including also all subsector integrals.

We compute the differential equations for the master integrals chosen through a generic integral ordering (see e.g.\ \cite{vonManteuffel:2012np}) by a straight-forward IBP reduction of their derivatives.
Here, we compute the derivatives both with respect to $t$ and $p^2$. 
In dimensional regularization, one can derive a relation between the partial derivatives from the behavior of the integral under a rescaling of all dimensionful parameters (see e.g.~\cite{Abreu:2022mfk}).
We observe that some of them are not manifestly fulfilled, which we interpret as a consequence of incomplete reductions due to missing ``hidden'' relations.
By enforcing these scaling relations, we obtain a number of additional identities, which relate integrals from different sectors with the same number of different propagators (and subsector integrals).
These additional relations simplify the differential equations in $t$ for the remaining master integrals and we cast them into $\epsilon$-form~\cite{Henn:2013pwa} using the codes \texttt{CANONICA}~\cite{Meyer:2016slj,Meyer:2017joq} and \texttt{Libra}~\cite{Lee:2014ioa,Lee:2020zfb}.
We obtain
\begin{equation}
    \mathrm{d}\vec{I}(t,\epsilon) = \epsilon \sum_i \mathrm{d}\ln(t-t_i) \mathbf{A}^{\!(i)} \vec{I}(t,\epsilon)\,,
\end{equation}
where we have set $p^2=-1$ and $\Delta \cdot p =1$. $\vec{I}$ is the vector of the new basis integrals, $\mathbf{A}^{\!(i)}$ are matrices of rational numbers and $t_i=0,\pm 1$.

Integrals without any $t$-dependent linear propagators are standard four-loop self-energy integrals~\cite{Baikov:2010hf,Lee:2011jt}, which in the present case were mapped to the planar two-point functions in \cite{vonManteuffel:2019gpr}.
We use these integrals to determine the purely $\epsilon$-dependent prefactors needed to transition from a basis in $\epsilon$-form to a basis with uniformly transcendental (UT) solutions.
In addition, they provide all the explicit boundary conditions that we need in addition to the following structural requirement.
Due to the fact that we introduce $t$ as a tracing parameter for a power series about $t=0$ and that our transformations of the integral coefficient are rational, the solutions of interest should have no branch cuts.
We take thus the limit $t \to 0$, solve the differential equation for exact $\epsilon$ and require the absence of branch cuts.
This gives additional conditions we can impose on our $\epsilon$-expanded solutions with exact $t$ dependence in the limit $t\to 0$.
In this way, we solve our master integrals as a Laurent expansion in $\epsilon$, where the coefficients are UT combinations of harmonic polylogarithms (HPLs)~\cite{Remiddi:1999ew} with weights $0,\pm 1$ and argument $t$ as well as zeta values.

We reduce the amplitude in terms of UT integrals and reconstruct the $d$ dependence as well as the $t$ dependence of the coefficients for a given finite field.
Here, we take advantage of the anticipated factorization of denominators and first determine the denominators \cite{Abreu:2018zmy,Heller:2021qkz} (as well as some simple overall numerator factors) in order to simplify the multivariate reconstruction.
We insert the $\epsilon$-expanded solutions for the master integrals into the amplitude and find the poles of the bare amplitude to contain transcendental functions of up to weight 7.
It is at this level of the $\epsilon$-expanded bare amplitude that we reconstruct the rational numbers from their image in a single finite field of cardinality $\mathcal{O}(2^{63})$.
A second finite field is used to check the reconstruction.

The result for the $N_f^2$ part of the bare four-loop OME $\braket{q|O^{\text{B}}_q|q}$ is then given in 
parameter-$t$ space. It is transformed to $n$-space in terms of harmonic sums~\cite{Vermaseren:1998uu,Blumlein:1998if} with the help of the package \texttt{HarmonicSums}~\cite{Ablinger:2009ovq,Ablinger:2012ufz,Ablinger:2014rba,Ablinger:2011te,Ablinger:2013cf,Ablinger:2014bra},
yielding the $n$-space expression for the bare 
OME.

\section{Renormalization and results}
\label{sec:results}
With the four-loop results for the bare OME computed in the above section in hand, all ingredients required for the renormalization procedure in~\eqref{eq:2qomeForOqRenor} are now available. We notice that the non-physical renormalization constant $Z_{qA}$ needs to be evaluated to three loops and the three-loop corrections of $\braket{q|O^{\text{B}}_q|q}$ need to be obtained to finite terms in $\epsilon$. Both of them were already obtained previously~\cite{Gehrmann:2023ksf}. Combining all components according to~\eqref{eq:2qomeForOqRenor}, the $N_f^2$ part of $Z_{qq}$ is determined to four-loop order and the corresponding anomalous dimension can be easily extracted from~\eqref{eq:ZfactorIntermsofGamma}. Similarly, the $N_f^2$ contributions for non-singlet anomalous dimensions are determined through~\eqref{eq:nsRenormalization} and \eqref{eq:ZfactorIntermsofGammaNS} and cross-checked against the results provided in~\cite{Davies:2016jie}. We do not repeat them here and restrict ourselves to displaying the new results for the pure-singlet anomalous dimension in $n$-space. We find:
\begin{align}
\gamma^{(3)}_{\text{ps}}&(n)\big|_{C_F N_f^3} =  \frac{16 \left(n^2+n+2\right)^2}{9 (n-1) n^2 (n+1)^2 (n+2)}  \left(6 \zeta _3-4 S_{1,1,1}\right)\nonumber 
   \\
   &+\frac{64 \left(8 n^7+37 n^6+83 n^5+85 n^4+61 n^3+58 n^2-20 n-24\right) S_{1,1}}{27 (n-1) n^3 (n+1)^3 (n+2)^2}\nonumber 
   \\
   &-\frac{32}{81 (n-1) n^4 (n+1)^4 (n+2)^3} \bigg[ 64 n^{10}+464
   n^9+1449 n^8+2037 n^7+1012 n^6 \nonumber 
   \\
   &  +105 n^5-411 n^4-1784 n^3-1304 n^2-48 n+144\bigg] S_1    +\frac{16}{81 (n-1) n^5 (n+1)^5 (n+2)^4} \bigg[\nonumber 
   \\
   & 64 n^{13}+576 n^{12}+2168 n^{11}+3376 n^{10}+258
   n^9-4248 n^8-3615 n^7-4520 n^6\nonumber 
   \\
   &-7225 n^5-1722 n^4+3304 n^3+1840 n^2-336 n-288  \bigg]\,,
   \label{eq:nf3psNspace}
\end{align}

\begin{align}
\gamma^{(3)}_{\text{ps}}&(n)\big|_{C_F^2 N_f^2} = \frac{1}{81 (n-1)^2 n^7 (n+1)^7 (n+2)^5} \bigg[5413 n^{19}+57482 n^{18}+254354 n^{17}\nonumber 
   \\
   & +639426 n^{16}+1396536 n^{15}+4120966 n^{14}+11962810 n^{13}+25755742 n^{12}\nonumber 
   \\
   &+38724195 n^{11}+32487128 n^{10}-11246124 n^9-77944456 n^8-127667936
   n^7\nonumber 
   \\
   &-142086912 n^6-123597440 n^5-83330816 n^4-43996416 n^3-18487296 n^2\nonumber 
   \\
   &-5557248 n-829440 \bigg]  -\frac{256 \left(n^2+n+2\right)^2 S_{-3}}{3 (n-1)^2 n^3 (n+1)^3 (n+2)^2} \nonumber 
   \\
   &-\frac{32
   \left(n^9+6 n^8+14 n^7+40 n^6+137 n^5+506 n^4+784 n^3+672 n^2+560 n+224\right) S_{-2}}{3 (n-1)^2 n^4 (n+1)^4 (n+2)^3}\nonumber 
   \\
   & - \frac{4}{81 (n-1) n^5 (n+1)^5 (n+2)^4} \bigg[6895 n^{13}+68895 n^{12}+296692 n^{11}+703266
   n^{10}\nonumber 
   \\
   &+1104101 n^9+1623673 n^8+2693912 n^7+3944838 n^6+4420952 n^5+3670248 n^4\nonumber 
   \\
   &+2067520 n^3+699296 n^2+91776 n-8064\bigg] S_1+\frac{8}{81 (n-1) n^4 (n+1)^4 (n+2)^2} \bigg[655 n^9\nonumber 
   \\
   &+4026
   n^8+8100 n^7-5502 n^6-38027 n^5-67940 n^4-80336 n^3-56704 n^2-20112 n\nonumber 
   \\
   &-2880\bigg] S_2\nonumber 
   \\
   &+\frac{32 \left(77 n^8+284 n^7+433 n^6-31 n^5-312 n^4-93 n^3-498
   n^2+260 n+456\right) S_3}{27 (n-1)^2 n^3 (n+1)^3 (n+2)^2}\nonumber 
   \\
   &-\frac{32 \left(23 n^8+86 n^7+217 n^6+461 n^5+270 n^4-99 n^3-18 n^2-52 n+264\right) \zeta _3}{9 (n-1)^2 n^3 (n+1)^3
   (n+2)^2}\nonumber 
   \\
   &+ \frac{8}{81 (n-1) n^4 (n+1)^4
   (n+2)^3}  \bigg[2501 n^{10}+18325 n^9+57780 n^8+96942 n^7\nonumber 
   \\
   &+120281 n^6+181533 n^5+246102 n^4+221096 n^3+156608 n^2+77328 n+19296\bigg] S_{1,1} \nonumber 
   \\
   & -\frac{16 \left(5 n^7+19 n^6-409 n^5-2675 n^4-4388 n^3-3896 n^2-2912 n-816\right) S_{1,2}}{27 (n-1) n^3 (n+1)^3 (n+2)^2}\nonumber 
   \\
   &-\frac{16 \left(91 n^7+425 n^6+661 n^5-721 n^4-2152
   n^3-2080 n^2-2224 n-816\right) S_{2,1}}{27 (n-1) n^3 (n+1)^3 (n+2)^2}\nonumber 
   \\
   &-\frac{8 \left(n^2+n+2\right) \left(133 n^4+226 n^3+457 n^2+284 n+4\right) S_{1,1,1}}{9 (n-1) n^3 (n+1)^3
   (n+2)}\nonumber 
   \\
   &+\frac{16 \left(n^2+n+2\right)^2}{9 (n-1) n^2 (n+1)^2 (n+2)}  \bigg(45 \zeta_4 -16 S_4-12 S_1 \zeta _3-8 S_{1,3}+2 S_{2,2}-4 S_{3,1}\nonumber 
   \\
   &-6 S_{1,1,2}+2 S_{1,2,1}+14 S_{2,1,1}+22 S_{1,1,1,1}\bigg) \,,
      \label{eq:nf2cf2psNspace}
\end{align}

\begin{align}
\gamma^{(3)}_{\text{ps}}&(n)\big|_{C_F C_A N_f^2} = -\frac{4}{81 (n-2) (n-1)^2 n^5 (n+1)^6 (n+2)^5} \bigg[7627 n^{17}+74941 n^{16}\nonumber 
   \\
   &+279109 n^{15}+298173 n^{14}-931399 n^{13}-3113931 n^{12}-2639893 n^{11}+1581811 n^{10}\nonumber 
   \\
   &+4213772 n^9+2908014 n^8+1843168 n^7+3691920 n^6+7684032
   n^5+11725312 n^4\nonumber 
   \\
   &+12278016 n^3+7778304 n^2+2654208 n+373248\bigg] \nonumber 
   \\
   &+\frac{16 \left(n^2+n+2\right) \left(5 n^2+5 n+6\right) S_{-4}}{3 (n-1) n^2 (n+1)^2
   (n+2)}\nonumber 
   \\
   &-\frac{16 \left(59 n^8+224 n^7+489 n^6+765 n^5+174 n^4-565 n^3-742 n^2-556 n-232\right) S_{-3}}{9 (n-1)^2 n^3 (n+1)^3 (n+2)^2}\nonumber 
   \\
   &+ \frac{8}{81 (n-2) (n-1)^2 n^4 (n+1)^4 (n+2)^3}  \bigg[209 n^{12}-389 n^{11}-380 n^{10}-10012
   n^9\nonumber 
   \\
   &-20569 n^8+24309 n^7+68716 n^6+129788 n^5+109304 n^4-126512 n^3-190400 n^2\nonumber 
   \\
   &-109632 n-33408\bigg] S_{-2}+\frac{4}{81 (n-1)^2 n^5 (n+1)^5
   (n+2)^4} \bigg[9353 n^{14}+87430
   n^{13}\nonumber 
   \\
   &+345549 n^{12}+674794 n^{11}+510973 n^{10}-273064 n^9-781035 n^8-890768 n^7\nonumber 
   \\
   &-1217312 n^6-1089848 n^5-309768 n^4+26496 n^3-132544 n^2-117888 n-21888\bigg] S_1 \nonumber 
   \\
   & + \frac{8}{81 (n-1)^2 n^4 (n+1)^4 (n+2)^3} \bigg[59 n^{11}-205 n^{10}+1091 n^9+6453 n^8+10481 n^7\nonumber 
   \\
   &+11791 n^6-4815 n^5-39715 n^4-37428 n^3-13424 n^2-2688 n-720\bigg] S_2\nonumber 
   \\
   &-\frac{8
   \left(367 n^8+1342 n^7+2387 n^6+1195 n^5-192 n^4-297 n^3-3870 n^2-380 n+1752\right) S_3}{27 (n-1)^2 n^3 (n+1)^3 (n+2)^2}\nonumber 
   \\
   &+\frac{16 \left(n^2+n+2\right) \left(23 n^2+23 n+34\right) S_4}{9
   (n-1) n^2 (n+1)^2 (n+2)}\nonumber 
   \\
   &+\frac{32 \left(7 n^8+29 n^7+69 n^6+147 n^5+94 n^4-36 n^3+18 n^2-32 n+88\right) \zeta _3}{3 (n-1)^2 n^3 (n+1)^3 (n+2)^2}\nonumber 
   \\
   &+\frac{16 \left(n^2+n+2\right) \left(23
   n^2+23 n+22\right) S_{-3,1}}{9 (n-1) n^2 (n+1)^2 (n+2)}-\frac{64 \left(n^2+n-10\right) \left(n^2+n+2\right) S_{-2,-2}}{9 (n-1) n^2 (n+1)^2 (n+2)}\nonumber 
   \\
   &+\frac{32 \left(71 n^7+367 n^6+866
   n^5+1609 n^4+1657 n^3+994 n^2+892 n+360\right) S_{-2,1}}{27 (n-1) n^3 (n+1)^3 (n+2)^2}\nonumber 
   \\
   &+\frac{32 \left(n^2+n+2\right) \left(7 n^2+7 n+2\right) S_{-2,2}}{9 (n-1) n^2 (n+1)^2
   (n+2)}+\frac{128 \left(n^2+n+1\right) \left(n^2+n+2\right) S_{1,-3}}{3 (n-1) n^2 (n+1)^2 (n+2)}\nonumber 
   \\
   &-\frac{64 \left(52 n^7+224 n^6+541 n^5+554 n^4+317 n^3+164 n^2-460 n-288\right) S_{1,-2}}{27
   (n-1) n^3 (n+1)^3 (n+2)^2}\nonumber 
   \\
   &-\frac{16}{81 (n-1)^2 n^4
   (n+1)^4 (n+2)^3} \bigg[1474 n^{11}+9886 n^{10}+28861 n^9+43998 n^8\nonumber 
   \\
   &+35410 n^7+22892 n^6+8871 n^5-36800 n^4-46032 n^3-3040 n^2+11664 n+2304\bigg] S_{1,1}\nonumber 
   \\
   & -\frac{16 \left(n^2+n+2\right) \left(40 n^5+138 n^4+325 n^3+471 n^2+190 n+24\right) S_{1,2}}{9 (n-1) n^3 (n+1)^3 (n+2)^2}\nonumber 
   \\
   &+\frac{16 \left(n^2+n+2\right) \left(5 n^2+5
   n+14\right) S_{1,3}}{3 (n-1) n^2 (n+1)^2 (n+2)}\nonumber 
   \\
   &+\frac{16 \left(n^2+n+2\right) \left(16 n^6+66 n^5-87 n^4-158 n^3-43 n^2-58 n-24\right) S_{2,1}}{27 (n-1)^2 n^3 (n+1)^3 (n+2)^2}\nonumber 
   \\
   &-\frac{32
   \left(n^2+n+2\right) \left(19 n^2+19 n-10\right) S_{-2,1,1}}{9 (n-1) n^2 (n+1)^2 (n+2)}-\frac{64 \left(n^2+n-10\right) \left(n^2+n+2\right) S_{1,-2,1}}{9 (n-1) n^2 (n+1)^2 (n+2)}\nonumber 
   \\
   &+\frac{32
   \left(n^2+n+2\right) \left(38 n^5+39 n^4+86 n^3+29 n^2-78 n-26\right) S_{1,1,1}}{9 (n-1)^2 n^3 (n+1)^3 (n+2)}\nonumber 
   \\
   &+\frac{16 \left(n^2+n+2\right)^2}{9 (n-1) n^2 (n+1)^2 (n+2)}  \bigg(-45 \zeta_4-6 S_{2,-2}-8 S_{2,2}+15 S_{3,1}+44
   S_{1,1,-2}\nonumber 
   \\
   &+29 S_{1,1,2}+15 S_{1,2,1}+2 S_{2,1,1}-22 S_{1,1,1,1}\bigg)\,,
      \label{eq:nf2cfcapsNspace}
\end{align}
where we omit the argument $n$ of the harmonic sums defined by 
\begin{align}
\label{eq:HarmonicDefinition}
S_{\pm m_1, \,m_2,\,\cdots m_d}(n) &= \sum_{j=1}^{n} (\pm 1)^{j} j^{-m_1} S_{m_2,\,\cdots m_d}(j) \quad(m_i \in \mathbb{N}),\nonumber\\
S_\emptyset(n)&=1\,.
\end{align}
The $N_f^3$ contribution in~\eqref{eq:nf3psNspace} was derived in~\cite{Gracey:1996ad,Davies:2016jie} and we find full agreement. The $N_f^2$ contributions with symbolic $n$ dependence in~\eqref{eq:nf2cf2psNspace} and~\eqref{eq:nf2cfcapsNspace} are new and constitute one of the main results in this paper. The all-$n$ coefficient of the $\zeta_4$ contribution was first predicted in \cite{Davies:2017hyl}, and we find full agreement upon correction of a typographical error according to~\cite{Falcioni:2023luc}. Evaluating our all-$n$ results for numerical $n$, we find full agreement with the fixed $n\leq 20$ results derived recently in~\cite{Falcioni:2023luc}. The anomalous dimensions are related to the splitting functions through the following Mellin transformation,
\begin{align}
\gamma_{ij}(n)  &= - \int_0^1 \,dx \, x^{n-1} \,P_{ij}(x).
\end{align} 
By an inverse Mellin transformation implemented in \texttt{HarmonicSums}, or by the method proposed in~\cite{Behring:2023rlq}, the above pure anomalous dimensions are transferred to the corresponding splitting functions:
\begin{align}
P_{\text{ps}}^{(3)}&(x)\big|_{C_F N_f^3} =- \frac{64}{27} (x-1) \left(4 x^2+7 x+4\right) \frac{1}{x} H_{1,1,1}+\frac{64}{27} \left(4 x^2-7 x-13\right) H_{2,1} \nonumber 
\\
&-\frac{64}{9} (x+1) \zeta _2 H_{0,0}-\frac{2336}{81} (x+1) H_{0,0}-\frac{64}{9} (x-1) (2 x-5)
   H_{1,1}-\frac{128}{9} (x+1) H_{3,1}\nonumber 
\\
&-\frac{464}{27} (x+1) H_{0,0,0}+\frac{128}{9} (x+1) H_{2,1,1}-\frac{32}{9} (x+1) H_{0,0,0,0}\nonumber 
\\
&+\frac{64}{81} \frac{1}{x}  (x-1) \left(34 x^2-49
   x-2\right)  H_1-\frac{64}{81} \frac{1}{x} (x-1) \left(7 x^2-12 x+3\right)\nonumber 
\\
&+\frac{64}{27} \frac{1}{x} \left(2 x^3-3 x^2-6 x-6\right) \zeta _3+\frac{32}{81}  \left(64 x^2-11
   x-59\right) H_0 -\frac{64}{81}  \left(38 x^2-x-49\right) H_2 \nonumber 
\\
& -\frac{928}{27} (x+1) \zeta _2  H_0 -\frac{128}{9} (x+1) \zeta _3  H_0+\frac{928}{27}  (x+1) H_3+\frac{64}{9}  (x+1)H_4\nonumber 
\\
&+\frac{64}{81}
   \left(38 x^2-x-49\right) \zeta _2-\frac{160}{9} (x+1) \zeta _4 \,,
       \label{eq:nf3psxspace}
\end{align}

\begin{align}
P^{(3)}_{\text{ps}}&(x)\big|_{C_F^2 N_f^2} =  \frac{ (x-1) \left(6632656 x^2+2333383 x+4127560\right)}{2187 x} \nonumber 
\\
&-\frac{2}{729} \frac{1}{x} \left(885688 x^3+1171023 x^2+473367 x+105088\right) H_0 \nonumber
\\
&-\frac{4}{243} \frac{1}{x} (x-1)
   \left(17492 x^2+52305 x+9746\right) H_1+\frac{8}{243} \left(4040 x^2+28557 x+28128\right) H_2 \nonumber
\\
&-\frac{8}{81} \left(4072 x^2-167 x-677\right) H_3+\frac{32}{27} \left(88 x^2+362 x+365\right)
   H_4+\frac{224}{9} (x+1) H_5  \nonumber
\\
&-\frac{8}{243} \frac{1}{x} \left(4040 x^3+48645 x^2+28128 x-9936\right) \zeta _2 \nonumber
\\
&+\frac{8}{81} \frac{1}{x} \left(4072 x^3+6421 x^2-677 x+736\right) H_0 \zeta
   _2 \nonumber
\\
&+\frac{16}{81} \frac{1}{x} (x-1) \left(884 x^2-403 x+308\right) H_1 \zeta _2  \nonumber
\\
& -\frac{64}{27} \left(46 x^2+49 x-14\right) H_2 \zeta _2+\frac{1088}{9} (x+1) H_3 \zeta _2 \nonumber
\\
&-\frac{64}{27}
   \frac{1}{x} \left(266 x^3-162 x^2+69 x-84\right) \zeta _3+\frac{32}{27} \frac{1}{x} \left(44 x^3+653 x^2+149 x-24\right) H_0 \zeta _3  \nonumber
\\
& -\frac{320}{9} (x+1) H_2 \zeta _3-\frac{1024}{9} (x+1)
   \zeta _2 \zeta _3+\frac{8}{27} \frac{1}{x} \left(1788 x^3+3746 x^2-151 x-288\right) \zeta _4  \nonumber
\\
&-\frac{8}{9} (743 x+959) H_0 \zeta _4-\frac{32}{9} (73 x+109) \zeta _5-\frac{128}{3} (3 x+7)
   H_{-4,0}  \nonumber
\\
&+\frac{128}{27} (x+3) (16 x-21) H_{-3,0}-\frac{32}{81} \frac{1}{x} \left(680 x^3-1647 x^2+2835 x-184\right) H_{-2,0}  \nonumber
\\
& +\frac{64}{9} \frac{1}{x} (x+1) \left(79 x^2-172 x+46\right)
   H_{-1,0}+\frac{32}{243} \left(8741 x^2+13596 x+19725\right) H_{0,0} \nonumber
\\
&-\frac{32}{27} \left(88 x^2+254 x+365\right) \zeta _2 H_{0,0}-\frac{128}{9} (19 x+10) \zeta _3 H_{0,0} \nonumber
\\
& +\frac{8}{243}
   \frac{1}{x} (x-1) \left(4484 x^2+2237 x+1910\right) H_{1,0} \nonumber
\\
& +\frac{16}{243} \frac{1}{x} (x-1) \left(496 x^2+12181 x+2971\right) H_{1,1}-\frac{16}{81} \frac{1}{x} (x-1) \left(884 x^2-403
   x+308\right) H_{1,2} \nonumber
\\
&-\frac{16}{81} \left(1852 x^2+995 x+878\right) H_{2,0}-\frac{128}{9} (x+1) \zeta _2 H_{2,0} \nonumber
\\
&-\frac{8}{81} \left(2616 x^2+3013 x+7771\right) H_{2,1} \nonumber
\\
&-\frac{64}{9} (x+1)
   \zeta _2 H_{2,1}+\frac{64}{27} \left(46 x^2+49 x-14\right) H_{2,2}+\frac{128}{9} (x+1) H_{2,3} \nonumber
\\
&+\frac{32}{27} \left(100 x^2+239 x+179\right) H_{3,0}+\frac{32}{27} \left(164 x^2+60
   x-99\right) H_{3,1}-\frac{1088}{9} (x+1) H_{3,2} \nonumber
\\
&-\frac{640}{9} (x+1) H_{4,0}-\frac{2144}{9} (x+1) H_{4,1}-\frac{256}{3} (x+1) H_{-3,0,0} \nonumber
\\
&+\frac{256}{27} \frac{1}{x} (x-1) \left(2 x^2+11
   x+2\right) H_{-2,0,0}-\frac{512}{81} \frac{1}{x} (x+1) \left(11 x^2+43 x+11\right) H_{-1,0,0} \nonumber
\\
&-\frac{8}{27} \left(3072 x^2-751 x-2259\right) H_{0,0,0}-\frac{32}{9} (43 x+7) \zeta _2
   H_{0,0,0} \nonumber
\\
&-\frac{32}{81} \frac{1}{x} (x-1) \left(200 x^2-853 x+56\right) H_{1,0,0}-\frac{16}{81} \frac{1}{x} (x-1) \left(1276 x^2+127 x+556\right) H_{1,1,0} \nonumber
\\
&-\frac{8}{27} \frac{1}{x} (x-1)
   \left(616 x^2+831 x+376\right) H_{1,1,1}+\frac{32}{27} \frac{1}{x} \left(64 x^3-91 x^2-163 x+8\right) H_{2,0,0} \nonumber
\\
&+\frac{64}{27} \left(54 x^2+104 x+29\right) H_{2,1,0}+\frac{32}{27} \left(44
   x^2+270 x+207\right) H_{2,1,1}+\frac{64}{9} (x+1) H_{2,1,2} \nonumber
\\
&-\frac{64}{9} (x+1) H_{2,2,0}+\frac{448}{9} (x+1) H_{2,2,1}-\frac{64}{9} (7 x+19) H_{3,0,0}-\frac{1088}{9} (x+1)
   H_{3,1,0} \nonumber
\\
&-\frac{32}{9} (x+1) H_{3,1,1}+\frac{32}{9} \left(32 x^2-167 x+476\right) H_{0,0,0,0}+\frac{1}{27} \frac{1}{x} (x-1) \left(4 x^2+7 x+4\right) \bigg[\nonumber
\\
&+160 H_1 \zeta _3+64 \zeta _2 H_{1,0} +32 \zeta _2 H_{1,1}-64 H_{1,3}-32
   H_{1,1,2}+32 H_{1,2,0}-224 H_{1,2,1}\nonumber
\\
&+256 H_{1,0,0,0}-128 H_{1,1,0,0}+96 H_{1,1,1,0}+352 H_{1,1,1,1}\bigg]-\frac{512}{9} (x+1) H_{2,0,0,0}\nonumber
\\
&+\frac{256}{9} (x+1) H_{2,1,0,0}-\frac{64}{3}
   (x+1) H_{2,1,1,0}-\frac{704}{9} (x+1) H_{2,1,1,1}\nonumber
\\
&-64 (x-4) H_{0,0,0,0,0}+320 (x+1) H_{0,0,0,0,0,0}   \,,     \label{eq:nf2cf2psxspace}
\end{align}

\begin{align}
P^{(3)}_{\text{ps}}&(x)\big|_{C_F C_A N_f^2} = -\frac{2  (x-1) \left(2687020 x^2+672220 x+725839\right)}{2187 x} \nonumber
\\
&+\frac{2}{729} \frac{1}{x} \left(929808 x^3+647385 x^2+501678 x+46688\right) H_0\nonumber
\\
&+\frac{2}{729} \frac{1}{x} (x-1)
   \left(268144 x^2+471253 x+161755\right) H_1\nonumber
\\
&-\frac{4}{243} \frac{1}{x} \left(20860 x^3+72921 x^2+79896 x+3296\right) H_2+\frac{8}{81} \left(1106 x^2-2935 x-2224\right) H_3\nonumber
\\
&+\frac{8}{27}
   \left(48 x^2-125 x-88\right) H_4-\frac{32}{9} (13 x-19) H_5\nonumber
\\
&+\frac{4}{243} \frac{1}{x} \left(20860 x^3+89265 x^2+79896 x-13014\right) \zeta _2+64 (x-3) H_{-3} \zeta _2\nonumber
\\
&+\frac{32}{27}
   \left(96 x^2+25 x-148\right) H_{-2} \zeta _2-\frac{32}{81} \frac{1}{x} (x+1) \left(868 x^2+623 x+418\right) H_{-1} \zeta _2\nonumber
\\
&-\frac{8}{81} \frac{1}{x} \left(1106 x^3+2575 x^2-2224
   x+368\right) H_0 \zeta _2\nonumber
\\
&+\frac{16}{81} \frac{1}{x} (x-1) \left(31 x^2+727 x+130\right) H_1 \zeta _2+\frac{16}{27} \frac{1}{x} \left(68 x^3-120 x^2-213 x-16\right) H_2 \zeta _2\nonumber
\\
&-\frac{64}{9}
   (13 x+11) H_3 \zeta _2+\frac{16}{81} \frac{1}{x} \left(3561 x^3+259 x^2-1172 x-180\right) \zeta _3-\frac{416}{3} (x-1) H_{-2} \zeta _3\nonumber
\\
&-\frac{16}{3} \frac{1}{x} (x+1) \left(8 x^2-35
   x+8\right) H_{-1} \zeta _3-\frac{32}{27} \frac{1}{x} \left(198 x^3+767 x^2+598 x-36\right) H_0 \zeta _3 \nonumber
\\
&-\frac{64}{27} \frac{1}{x} (x-1) \left(65 x^2+161 x+65\right) H_1 \zeta
   _3+\frac{2752}{9} (x+1) H_2 \zeta _3-\frac{32}{9} (59 x+23) \zeta _2 \zeta _3\nonumber
\\
&-\frac{2}{27} \frac{1}{x} \left(3560 x^3+12229 x^2+4664 x-2192\right) \zeta _4+\frac{16}{9} (157 x+178) H_0
   \zeta _4\nonumber
\\
&+\frac{32}{9} (230 x+9) \zeta _5+\frac{64}{9} (9 x+29) H_{-4,0}-\frac{16}{27} \left(216 x^2+99 x+98\right) H_{-3,0}\nonumber
\\
&-\frac{64}{3} (x-5) H_{-3,2}+\frac{448}{3} (x-1) \zeta _2
   H_{-2,-1} \nonumber
\\
& +\frac{16}{81} \frac{1}{x} \left(2560 x^3-2755 x^2+4000 x-184\right) H_{-2,0}-\frac{64}{9} (x-1) \zeta _2 H_{-2,0} \nonumber
\\
&-\frac{32}{27} \left(56 x^2-19 x-47\right) H_{-2,2}-\frac{352}{9}
   (x-1) H_{-2,3} \nonumber
\\
& +\frac{32}{9} \frac{1}{x} (x+1) \left(20 x^2-53 x+20\right) \zeta _2 H_{-1,-1}\nonumber
\\
&-\frac{8}{243} \frac{1}{x}^2 (x+1) \left(19756 x^3-27928 x^2+8131 x+24\right)
   H_{-1,0}\nonumber
\\
&+\frac{32}{27} \frac{1}{x} (x+1) \left(20 x^2+19 x+20\right) \zeta _2 H_{-1,0}+\frac{32}{81} \frac{1}{x} (x+1) \left(584 x^2+277 x+332\right) H_{-1,2}\nonumber
\\
&-\frac{16}{27} \frac{1}{x} (x+1)
   \left(68 x^2-65 x+68\right) H_{-1,3}-\frac{8}{81} \left(9108 x^2+10984 x+3451\right) H_{0,0}\nonumber
\\
&-\frac{8}{27} \left(48 x^2+73 x-88\right) \zeta _2 H_{0,0}+\frac{32}{9} (121 x+23) \zeta _3
   H_{0,0}\nonumber
\\
&+\frac{4}{243} \frac{1}{x} (x-1) \left(9184 x^2-3725 x-2093\right) H_{1,0}-\frac{128}{27} \frac{1}{x} (x-1) \left(11 x^2+26 x+11\right) \zeta _2 H_{1,0}\nonumber
\\
&+\frac{16}{243} \frac{1}{x}
   (x-1) \left(1949 x^2-13441 x-2218\right) H_{1,1}-\frac{16}{81} \frac{1}{x} (x-1) \left(599 x^2+35 x+302\right) H_{1,2}\nonumber
\\
&-\frac{8}{81} \frac{1}{x} \left(242 x^3+911 x^2-1357 x-160\right)
   H_{2,0}+\frac{896}{9} (x+1) \zeta _2 H_{2,0}\nonumber
\\
&-\frac{16}{81} \frac{1}{x} \left(366 x^3-2633 x^2-2891 x-80\right) H_{2,1}+\frac{224}{9} (x+1) \zeta _2 H_{2,1}\nonumber
\\
&+\frac{16}{27} \frac{1}{x}
   \left(12 x^3+32 x^2+11 x+16\right) H_{2,2}-\frac{160}{3} (x+1) H_{2,3}+\frac{8}{27} \left(16 x^2+134 x-7\right) H_{3,0}\nonumber
\\
&+\frac{32}{27} \left(8 x^2+66 x+75\right) H_{3,1}+\frac{64}{9} (7
   x-1) H_{3,2}-\frac{16}{9} (23 x-5) H_{4,0}+\frac{128}{9} (x-4) H_{4,1}\nonumber
\\
&+\frac{256}{3} (x-2) H_{-3,-1,0}+\frac{32}{9} (3 x+29) H_{-3,0,0}-\frac{64}{3} (x-1) H_{-2,-2,0}\nonumber
\\
&+\frac{64}{27}
   \left(40 x^2+44 x-101\right) H_{-2,-1,0}-\frac{640}{9} (x-1) H_{-2,-1,2}\nonumber
\\
&-\frac{16}{27} \frac{1}{x} \left(136 x^3-213 x^2+69 x-16\right) H_{-2,0,0}-\frac{64}{9} (x-1)
   H_{-2,2,0}+\frac{320}{9} (x-1) H_{-2,2,1}\nonumber
\\
&-\frac{32}{9} \frac{1}{x} (x+1) \left(4 x^2-7 x+4\right) H_{-1,-2,0}-\frac{128}{81} \frac{1}{x} (x+1) \left(142 x^2+173 x+43\right)
   H_{-1,-1,0}\nonumber
\\
&-\frac{64}{27} \frac{1}{x} (x+1) \left(8 x^2-41 x+8\right) H_{-1,-1,2}+\frac{16}{81} \frac{1}{x} (x+1) \left(1112 x^2+679 x+752\right) H_{-1,0,0}\nonumber
\\
&-\frac{32}{27} \frac{1}{x} (x+1)
   \left(16 x^2-x+16\right) H_{-1,2,0}-\frac{32}{27} \frac{1}{x} (x+1) \left(28 x^2+59 x+28\right) H_{-1,2,1}\nonumber
\\
&-\frac{8}{81} \left(8 x^2+3121 x-2472\right) H_{0,0,0}+\frac{32}{9} (31 x-19)
   \zeta _2 H_{0,0,0}\nonumber
\\
&-\frac{64}{27} \frac{1}{x} (x-1) \left(8 x^2+41 x+8\right) H_{1,-2,0}+\frac{8}{27} \frac{1}{x} (x-1) \left(136 x^2-1455 x-8\right) H_{1,0,0}\nonumber
\\
&+\frac{16}{27} \frac{1}{x} (x-1)
   \left(85 x^2-3 x+82\right) H_{1,1,0}+\frac{32}{81} \frac{1}{x} (x-1) \left(385 x^2+481 x+205\right) H_{1,1,1}\nonumber
\\
&+\frac{640}{9} (x+1) H_{2,-2,0}-\frac{8}{27} \frac{1}{x} \left(176 x^3-791
   x^2-971 x+32\right) H_{2,0,0}\nonumber
\\
&-\frac{16}{27} \left(44 x^2+240 x+195\right) H_{2,1,0}-\frac{128}{27} \frac{1}{x} \left(11 x^3+63 x^2+39 x-11\right) H_{2,1,1}\nonumber
\\
&+\frac{160}{3} (x+1)
   H_{2,1,2}+\frac{256}{9} (x+1) H_{2,2,0}+\frac{64}{9} (x+1) H_{2,2,1}+\frac{16}{9} (17 x+73) H_{3,0,0}\nonumber
\\
&-\frac{64}{9} (5 x-3) H_{3,1,0}+\frac{64}{9} (10 x+13) H_{3,1,1}+\frac{1408}{9} (x-1)
   H_{-2,-1,-1,0}\nonumber
\\
&+\frac{128}{3} (x-1) H_{-2,-1,0,0}+32 (x-1) H_{-2,0,0,0}+\frac{704}{27} \frac{1}{x} (x+1) \left(4 x^2-7 x+4\right) H_{-1,-1,-1,0}\nonumber
\\
&+\frac{128}{3} \frac{1}{x} (x+1)
   \left(x^2-x+1\right) H_{-1,-1,0,0}+\frac{16}{9} \frac{1}{x} (x+1) \left(16 x^2-19 x+16\right) H_{-1,0,0,0}\nonumber
\\
&+\frac{16}{27} (551 x-486) H_{0,0,0,0}-\frac{16}{27} \frac{1}{x} (x-1) \left(80
   x^2+113 x+80\right) H_{1,0,0,0}\nonumber
\\
&+\frac{16}{3}\frac{1}{x} (x-1) \left(8 x^2+17 x+8\right) H_{1,1,0,0}+\frac{1}{27} (x-1) \frac{1}{x} \left(4 x^2+7 x+4\right) \bigg[\nonumber
\\
&-112 \zeta _2 H_{1,1}+240 H_{1,3}-240 H_{1,1,2}-128 H_{1,2,0}-32
   H_{1,2,1}-464 H_{1,1,1,0}\nonumber
\\
&-352 H_{1,1,1,1}\bigg]+\frac{544}{9} (x+1) H_{2,0,0,0}-\frac{224}{3} (x+1) H_{2,1,0,0}\nonumber
\\
&+\frac{928}{9} (x+1) H_{2,1,1,0}+\frac{704}{9} (x+1)
   H_{2,1,1,1}+\frac{448}{9} x H_{0,0,0,0,0}\,. 
     \label{eq:nf2cfcapsxspace}
\end{align}
Here, we use the symbol $H$ to denote harmonic polylogarithms (HPLs) and omit the argument $x$. The HPLs are defined recursively by 
\begin{align}
H_{a_1,\,a_2,\,\cdots,\, a_m}(x) &= \int_{0}^x \ud t\, f_{a_1}(t) H_{a_2,\,\cdots,\,a_m}(t)\,,\nonumber\\
H_{\vec{0}_m}(x) &= \frac{\log^m x}{m!}\,,\nonumber\\
H(x)&=1\,,
\end{align}
where $a_i$ is $0$, $1$, or $-1$, and the kernel $f_{a}(t)$ is defined as 
\begin{align}
f_1(t) = \frac{1}{1-t}\,,\quad f_0(t) = \frac{1}{t}\,, \quad f_{-1}(t) = \frac{1}{1+t}\,.
\end{align}
In the above, we  adopt an abbreviated notation proposed in~\cite{Remiddi:1999ew,Maitre:2005uu}, for example 
\begin{align}
    H_{0,-1,0,0,1} = H_{-2,3};\qquad 
    H_{1,0,1,0} = H_{1,2,0}.
\end{align}
Our result for $N_f^3$ shown in~\eqref{eq:nf3psxspace} validated the corresponding result presented in~\cite{Davies:2016jie}. The $N_f^2$ contributions shown in~\eqref{eq:nf2cf2psxspace} and~\eqref{eq:nf2cfcapsxspace} are presented for the first time. We observe that these contributions are expressed through functions up to transcendental weight 6. Interestingly, the only transcendental function of the highest weight is $\log^6 x$.

With the analytic results in $x$-space, it is easy to extract the limit around $x = 0$ to high powers. For simplicity, we only present the results to the next-to-leading power,
\begin{align}
P^{(3)}_{\text{ps}}&(x)\big|_{C_F N_f^3} =  \frac{1}{x} \bigg[
   \frac{64}{27}-\frac{128 \zeta _3}{9}   \bigg] + \left(-\frac{32 \zeta _2}{9}-\frac{1168}{81}\right) \log ^2(x) \nonumber
  \\
  &+\left(-\frac{928 \zeta _2}{27}-\frac{128 \zeta _3}{9}-\frac{1888}{81}\right)
   \log (x)-\frac{4}{27} \log^4(x)-\frac{232 \log ^3(x)}{81} \nonumber
  \\
  &-\frac{3136 \zeta _2}{81}-\frac{128 \zeta _3}{9}-\frac{160 \zeta
   _4}{9}-\frac{832}{81} + \mathcal{O}(x) \,,
   \\
  P^{(3)}_{\text{ps}}&(x)\big|_{C_F^2 N_f^2} =  \frac{1}{x} \bigg[ \left(\frac{5888 \zeta
   _2}{81}-\frac{256 \zeta _3}{9}-\frac{210176}{729}\right) \log (x)+\frac{2944 \zeta _2}{9} \nonumber
   \\
   &+\frac{1792 \zeta _3}{9}-\frac{256 \zeta
   _4}{3}-\frac{4127560}{2187}  \bigg] +\left(\frac{8524}{81}-\frac{112 \zeta _2}{27}\right) \log ^3(x) \nonumber
   \\
   &+\left(-\frac{5840 \zeta _2}{27}-\frac{640 \zeta
   _3}{9}+\frac{104432}{81}\right) \log ^2(x) \nonumber
   \\
   &+\left(-\frac{6184 \zeta _2}{81}+\frac{4768 \zeta _3}{27}-\frac{7672 \zeta
   _4}{9}-\frac{75598}{81}\right) \log (x)+\frac{4 \log ^6(x)}{9}+\frac{32 \log ^5(x)}{15} \nonumber
   \\
   &+\frac{1904 \log ^4(x)}{27}-\frac{79168 \zeta
   _2}{81}-\frac{1024 \zeta _2 \zeta _3}{9}-\frac{5056 \zeta _3}{27}-\frac{1208 \zeta _4}{27}-\frac{3488 \zeta _5}{9}+\frac{43195}{81} + \mathcal{O}(x)\,,
    \\
  P^{(3)}_{\text{ps}}&(x)\big|_{C_F C_A N_f^2} =  \frac{1}{x} \bigg[ \left(-\frac{2944 \zeta _2}{81}+\frac{128 \zeta _3}{3}+\frac{92800}{729}\right) \log (x)-\frac{1928 \zeta
   _2}{9}-\frac{320 \zeta _3}{9}\nonumber
   \\
   &+\frac{4384 \zeta _4}{27}+\frac{1453406}{2187} \bigg] + \left(\frac{160}{3}-\frac{304 \zeta _2}{27}\right) \log ^3(x)+\left(\frac{352 \zeta _2}{27}+\frac{368 \zeta _3}{9}-\frac{1196}{9}\right)
   \log ^2(x)\nonumber
   \\
   &+\left(\frac{23936 \zeta _2}{81}-\frac{19136 \zeta _3}{27}+\frac{2848 \zeta _4}{9}+\frac{9424}{9}\right) \log (x) -12 \log
   ^4(x)+\frac{84160 \zeta _2}{81}\nonumber
   \\
   &-\frac{736 \zeta _2 \zeta _3}{9}-\frac{9728 \zeta _3}{81}-\frac{9328 \zeta _4}{27}+32 \zeta
   _5-\frac{15964}{81} + \mathcal{O}(x) \,.
\end{align}
    The terms $\log^2 x/x$ at leading power for $P^{(3)}_{\text{ps}}$ have been predicted some time ago in~\cite{Catani:1994sq} and they vanish for the $N_f^2$ contributions. The terms $\log^k x$ with $k=6\,, 5\,, 4$ at sub-leading power were predicted in~\cite{Davies:2022ofz}. Our results are consistent with~\cite{Catani:1994sq,Davies:2022ofz} and provide extra information that may be helpful to generalize the frameworks in~\cite{Catani:1994sq,Davies:2022ofz}.

In the limit of $x \to 1$, the pure-singlet splitting function is power-suppressed~\cite{Soar:2009yh}, we only show the result to the lowest power (proportional to $1-x$),
\begin{align}
  P^{(3)}_{\text{ps}}&(x)\big|_{C_F N_f^3} = (1-x)\bigg[  -\frac{32}{27} \log ^3(1-x)-\frac{160}{27} \log ^2(1-x)\nonumber
    \\
    &-\frac{1088}{81} \log (1-x)-\frac{32 \zeta _3}{3}+\frac{64}{81} \bigg]+ \mathcal{O}\left((1-x)^2\right)\,, 
    \\
    P^{(3)}_{\text{ps}}&(x)\big|_{C_F^2 N_f^2} = (1-x) \bigg[ \left(\frac{16 \zeta _2}{3}-\frac{6800}{81}\right) \log ^2(1-x) \nonumber
    \\
    &+\left(-\frac{368 \zeta _2}{27}-\frac{160 \zeta _3}{9}-\frac{4660}{27}\right) \log (1-x)-\frac{44}{27} \log
   ^4(1-x)-\frac{356}{27} \log ^3(1-x) \nonumber
    \\
    &-\frac{4136 \zeta _2}{81}+\frac{3056 \zeta _3}{27}-\frac{952 \zeta _4}{9}+\frac{8567}{81} \bigg] + \mathcal{O}\left((1-x)^2\right)\,, 
    \\
   P^{(3)}_{\text{ps}}&(x)\big|_{C_F C_A N_f^2}= (1-x)\bigg[ \left(\frac{7904}{81}-\frac{56 \zeta _2}{9}\right) \log ^2(1-x)\nonumber
    \\
    &+\left(\frac{80 \zeta _2}{27}-\frac{80 \zeta _3}{9}+\frac{21604}{81}\right) \log (1-x)+\frac{44}{27} \log ^4(1-x)+16 \log
   ^3(1-x)\nonumber
    \\
    &+\frac{124 \zeta _2}{81}-\frac{2392 \zeta _3}{27}+\frac{236 \zeta _4}{3}+\frac{2968}{27} \bigg] + \mathcal{O}\left((1-x)^2\right)\,.
\end{align}  
For the limit $x\to 1$, the terms $(1-x)^j \log(1-x)^k$ with $k=3,4$ have been predicted for all $j$ in~\cite{Soar:2009yh}. Their results for $j=1$ agree with our results shown above and we also provide the previously unknown contributions with $k=2,1,0$.

It is interesting to compare our analytic results in $x$-space with an approximation based on all previously known information for the $N_f^2$ contribution to $P^{(3)}_{\text{ps}}$. 
We determine this approximation by closely following the methodology outlined in~\cite{Falcioni:2023luc}, adopting the two representative functional forms $A$ and $B$ that were introduced therein,
\begin{align}
    \label{eq:fitA}
    P_{\text{ps},\,A}^{(3)}(x)\big|_{N_f^2} =& p_{\text{ps},\,0}(x)\big|_{N_f^2} + \sum_{i=1}^8 a_i~ f_i(x)  + a_9~(1-x)(1+2 x) + a_{10}~(1-x) x^2 \,, \\ 
    \label{eq:fitB}
    P_{\text{ps},\,B}^{(3)}(x)\big|_{N_f^2} =& p_{\text{ps},\,0}(x)\big|_{N_f^2} + \sum_{i=1}^8 b_i~ f_i(x)  + b_9~(1-x) + b_{10}~(1-x) x(1+ x) \,, \\
    f_i = \Big\{(1-x) &\frac{\log(x)}{x} ,~  \frac{1-x}{x} ,~  \log^3(x),~  \log^2(x) ,~ (1-x) \log(x) ,~\\ \nonumber
        & \hspace{5ex} (1-x) \log^2(1-x) ,~  (1-x) \log(1-x) ,~  (1-x)^2 \log^2(1-x)\Big\}
\end{align}
where $p_{\text{ps},\,0}\big|_{N_f^2}$ contains the terms $\{\log^6 (x),~\log^5 (x),~\log^4 (x),~(1-x) \log^3(1-x),~(1-x)^2 \log^3(1-x),~\log^4(1-x)\}$ that are predicted in~\cite{Davies:2022ofz,Soar:2009yh}.
The coefficients $\{a_i\}$, $\{b_i\}$ are fitted using the fixed $n \leq 20$ result.
The average of the two fits is then used as the central prediction, and the spread between as an uncertainty estimate. 
We furthermore check the consistency of our approximation by extracting 
the $N_f^2$ coefficient from the approximations presented in~\cite{Falcioni:2023luc} with fixed $N_f=3,\,4,\,5$, using the ansatz: $P^{(3)}_{\text{ps}}  = c_1 N_f + c_2 N_f^2 + c_3 N_f^3$, finding good agreement with  
 our fitted approximation described above. 
 
\begin{figure}[h]
  \centering
  \includegraphics[width=0.8\textwidth]{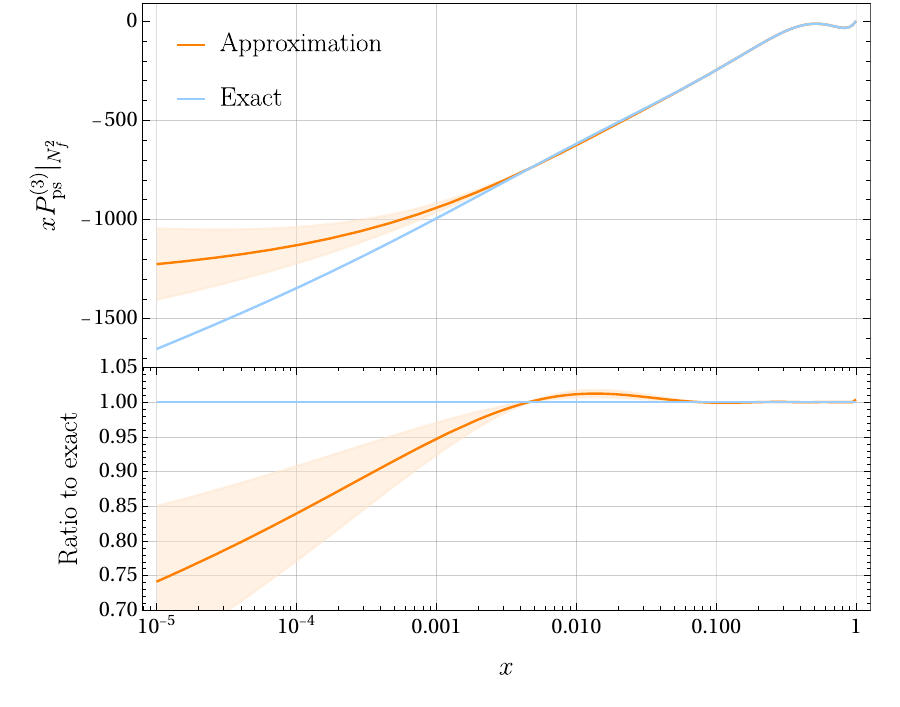}
  \caption{
    The $N_f^2$ contribution to the four-loop pure-singlet splitting function $P^{(3)}_{\text{ps}}$.
    The exact result derived in this work (blue) and an approximation based on $n\leq 20$ fixed moments and known $x\to0,1$ terms (orange) are compared.
The methodology to determine the approximation and its uncertainty are explained in the text. The bottom panel shows the ratio to the exact result.
  }
  \label{fig:fit-vs-exact}
\end{figure}

In Figure~\ref{fig:fit-vs-exact} we compare the approximation with our exact result. We observe that the approximation describes the exact result well in the large and moderate-$x$ region, with the deviations being below at most a few percent.
However, the small-$x$ region below $x\lesssim 10^{-3}$ is not captured correctly by the approximation. 
Its uncertainty is underestimated and the best-fit values are systematically above the exact result.
This highlights the importance of the divergent terms in the $x\to 0$ limit which were previously unknown and are difficult to constrain from a limited number of Mellin moments.

It is worth noting that our exact results allow one to construct a simple power-logarithmic approximation (involving only the terms of the form $x^j \log^k(x),~ (1-x)^j \log^k(1-x)$)
that is precise over the whole $x$ range. For the readers' convenience, we provide such an approximation below:
\begin{align}
 P^{(3)}_{\text{ps}}&(x)\big|_{N_f^3} = -\frac{19.6341}{x}-0.197531 \log ^4(x)-3.81893 \log ^3(x)-27.0245 \log ^2(x)\nonumber 
 \\ 
  &-129.255 \log (x)-147.059  -0.660017 x^5+0.752343 x^4-34.9402 x^3\nonumber 
 \\ 
  &+x^2 \Big[0.252007 \log ^4(x)+0.378676 \log ^3(x)+8.79762 \log ^2(x)+32.774 \log (x)\nonumber 
 \\ 
  &+106.766\Big]+(1-x) \Big[-1.57998 \log ^3(1-x)-7.89567 \log
   ^2(1-x)\nonumber 
 \\ 
  &-17.8746 \log (1-x)\Big]+(1-x)^2 \Big[-0.185609 \log ^3(1-x)-1.90667 \log ^2(1-x)\nonumber 
 \\ 
  &-7.06942 \log (1-x)\Big] +x \Big[-0.198775 \log ^4(x)-3.87072 \log ^3(x)-27.8681 \log
   ^2(x)\nonumber 
 \\ 
  &-110.358 \log (x)+94.7747\Big] \,,  \\
  P^{(3)}_{\text{ps}}&(x)\big|_{N_f^2}=  \frac{114.438 \log (x)}{x} -\frac{356.614}{x}+0.790123 \log ^6(x)+3.79259 \log ^5(x) \nonumber \\
  & +77.3663 \log ^4(x)+314.204 \log ^3(x)+1258.40 \log ^2(x)+949.678 \log (x)-49.9323 \nonumber \\
  & +  33.627 x^5-287.064 x^4+9680.22 x^3 +x^2 \Big[-4.63175 \log ^6(x)-11.7058 \log ^5(x)\nonumber \\
  &-393.364 \log ^4(x)-418.965 \log ^3(x)-8435.97 \log ^2(x)+1319.27 \log (x)-21652.5\Big]\nonumber \\
  &+(1-x)^2
   \Big[-4.84858 \log ^4(1-x)+8.27339 \log ^3(1-x)-92.6823 \log ^2(1-x)\nonumber \\
  &+161.418 \log (1-x)\Big]+(1-x) \Big[3.63945 \log ^4(1-x)+41.3407 \log ^3(1-x) \nonumber \\
  & +228.914 \log ^2(1-x)+761.999 \log
   (1-x)\Big]+x \Big[ 0.783717 \log ^6(x)+0.335522 \log ^5(x)\nonumber \\
  &+1.53492 \log ^4(x)-192.192 \log ^3(x)+358.62 \log ^2(x)-884.842 \log (x)+12632.2\Big]\nonumber \\
  &+0.0000703741 \,,
\end{align}
where we have set $C_A = 3,\, C_F = 4/3$ and truncated the numerical values of the coefficients to 6 digits. This approximation has an accuracy better than $10^{-5}$ over the whole range of $x$. It can be readily included in the programs implementing scale evolution of parton distribution functions.
\FloatBarrier

\section{Conclusions and Outlook}
\label{sec:outlook}
In this paper, we derived the renormalization of the quark singlet operator to four-loop order.
We observe that it does not require the computation of new renormalization counterterms beyond those that 
were already obtained for symbolic Mellin-$n$ in~\cite{Gehrmann:2023ksf}. 

As a first non-trivial application, we computed the $N_f^2$ contributions to the four-loop pure-singlet splitting functions. Our workflow to calculate the relevant four-loop OMEs is described in detail and we validate it on an independent rederivation of the four-loop $N_f^2$ contributions 
to the non-singlet splitting functions, giving full agreement with~\cite{Davies:2016jie}.
We employ our setup to derive the pure-singlet contributions involving two closed fermion loops, for the first time for symbolic $n$.
This allowed us to derive the exact results in $x$-space and to perform a comparison with an approximation obtained from the fixed $n \leq 20$ results.
We demonstrated that the approximation is adequate for large to moderate values of $x$, but fails to correctly capture the behavior at small values of $x$.
It would be interesting to investigate if this has tangible implications for the construction of approximate N\textsuperscript{3}LO parton distributions functions~\cite{McGowan:2022nag,Hekhorn:2023gul}.

Building on the methodology presented here, the complete computation of all color and flavor structures for the singlet, four-loop splitting functions can be envisaged. Towards this objective, we expect a considerable increase in complexity, in particular for the required integral reductions. Moreover, the computation of the splittings into gluons involves the renormalization of the 
gluon operator up to four loops and will likely require new counterterms whose Feynman rules remain to be determined. 

\begin{acknowledgments}
We would like to thank Giulio Falcioni, Franz Herzog, Sven-Olaf Moch and Andreas Vogt for constructive discussions. We acknowledge the European Research Council (ERC) for funding of this work under the European Union's Horizon 2020 research and innovation programme grant agreement 101019620 (ERC Advanced Grant TOPUP) and the National Science Foundation (NSF) for support under grant number 2013859.
\end{acknowledgments}

\appendix

\section{Standard QCD renormalization constant}
\label{app:qcdRenor}
The strong coupling renormalization constant $Z_{a_s}$ can be expressed through the QCD $\beta$ function,
\begin{align}
    Z_{a_s} = 1 - \frac{\beta_0}{\epsilon} a_s + \left(  \frac{\beta_0^2}{\epsilon^2} - \frac{\beta_1}{2 \epsilon}\right) a_s^2 + \left( -\frac{\beta_0^3}{\epsilon^3} + \frac{7 \beta_0 \beta_1}{6 \epsilon^2} - \frac{\beta_2}{ 3 \epsilon} \right) a_s^3 + \mathcal{O}(a_s^4).
\end{align}
Up to three loops, the QCD beta function reads \cite{Tarasov:1980au}
\begin{align}
  \beta_0& = \frac{11 C_A}{3} - \frac{2 N_f}{3} \,,  
  \\
    \beta_1 &= -\frac{10 C_A N_f}{3}+\frac{34 C_A^2}{3}-2 C_F N_f \,, \\
    \beta_2 &= -\frac{1415}{54} C_A^2 N_f+\frac{79}{54} C_A N_f^2-\frac{205}{18} C_A C_F N_f+\frac{2857 C_A^3}{54}+\frac{11}{9} C_F N_f^2+C_F^2 N_f \,.
\end{align}

The gluon field renormalization constant
\begin{align}
    Z_g = \sum_{l=0}^\infty Z_g^{(l)} a_s^l
\end{align}
is required to three-loop order~\cite{Tarasov:1980au,Larin:1993tp} for this work.
The relevant expansion coefficients read
\begin{align}
 Z_g^{(0)}  = &1 \,, 
     \\
    Z_g^{(1)}  = &\frac{1}{\epsilon}
   \bigg[ \left(\frac{13}{6}-\frac{\xi }{2}\right) C_A-\frac{2 N_f}{3}\bigg] \,, 
   \\
     Z_g^{(2)}  =  &\frac{1}{\epsilon} \bigg[-\frac{5 C_A N_f}{4}+\left(-\frac{\xi
   ^2}{8}-\frac{11 \xi }{16}+\frac{59}{16}\right) C_A^2-C_F
   N_f\bigg]\nonumber 
   \\
   & +\frac{1}{\epsilon^2} \bigg[\left(\frac{\xi }{3}+\frac{1}{2}\right) C_A
   N_f+\left(\frac{\xi ^2}{4}-\frac{17 \xi }{24}-\frac{13}{8}\right)
   C_A^2\bigg] \,, 
   \\
    Z_g^{(3)}  =  & \frac{1}{\epsilon^2} \bigg[C_F \bigg\{\left(\frac{\xi
   }{2}+\frac{31}{18}\right) C_A N_f-\frac{2 N_f^2}{9}\bigg\}+\left(\frac{\xi
   ^2}{12}+\frac{19 \xi }{24}+\frac{481}{108}\right) C_A^2 N_f-\frac{25}{54} C_A
   N_f^2\nonumber 
   \\
   &+\left(\frac{7 \xi ^3}{48}+\frac{13 \xi ^2}{24}-\frac{143 \xi
   }{96}-\frac{7957}{864}\right) C_A^3\bigg]+\frac{1}{\epsilon} \bigg[C_F \bigg\{\left(-4
   \zeta _3-\frac{5}{108}\right) C_A N_f+\frac{11 N_f^2}{27}\bigg\}\nonumber 
   \\
   & +C_A^2 N_f
   \left(\frac{\xi }{3}+3 \zeta _3-\frac{911}{108}\right)+\frac{19}{27} C_A
   N_f^2+C_A^3 \bigg\{-\frac{7 \xi ^3}{96}+\xi ^2 \left(-\frac{\zeta
   _3}{16}-\frac{11}{32}\right)\nonumber 
   \\
   &+\xi  \left(-\frac{\zeta
   _3}{4}-\frac{167}{96}\right)-\frac{3 \zeta
   _3}{16}+\frac{9965}{864}\bigg\}+\frac{1}{3} C_F^2 N_f\bigg]+\frac{1}{\epsilon^3}
   \bigg[\left(-\frac{\xi ^2}{6}-\frac{5 \xi }{12}-\frac{11}{9}\right) C_A^2
   N_f\nonumber 
   \\
   &+\frac{1}{9} C_A N_f^2+\left(-\frac{\xi ^3}{8}+\frac{\xi ^2}{6}+\frac{47 \xi
   }{48}+\frac{403}{144}\right) C_A^3\bigg] \,.
\end{align}

The quark field renormalization constant
\begin{align}
    Z_q = \sum_{l=0}^\infty Z_q^{(l)} a_s^l
\end{align}
is required to four-loop order~\cite{Chetyrkin:1999pq,Chetyrkin:2004mf}.
The relevant expansion coefficients read
\begin{align}
 Z_q^{(0)}  = &1 \,, 
     \\
     Z_q^{(1)} = &- C_F \frac{\xi}{\epsilon}  \,, 
    \\
      {Z_q^{(2)}} = &  \frac{C_F}{\epsilon} \bigg[\left(-\frac{\xi ^2}{8}-\xi -\frac{25}{8}\right)
   C_A+\frac{3 C_F}{4}+\frac{N_f}{2}\bigg]+\frac{C_F}{\epsilon^2} \bigg[\left(\frac{\xi
   ^2}{4}+\frac{3 \xi }{4}\right) C_A+\frac{\xi ^2 C_F}{2}\bigg] \,, 
   \\
    {Z_q^{(3)}} = & \frac{C_F}{\epsilon^3} \bigg[C_A \left(\Big(-\frac{\xi ^3}{4}-\frac{3 \xi
   ^2}{4}\Big) C_F+\frac{\xi  N_f}{6}\right)+\left(-\frac{\xi ^3}{12}-\frac{3 \xi
   ^2}{8}-\frac{31 \xi }{24}\right) C_A^2-\frac{1}{6} \xi ^3
   C_F^2\bigg] \nonumber 
   \\
   &+\frac{C_F}{\epsilon^2} \bigg[C_A \left(\left(\frac{\xi ^3}{8}+\xi
   ^2+\frac{25 \xi }{8}-\frac{11}{6}\right) C_F+\left(-\frac{\xi
   }{2}-\frac{47}{18}\right) N_f\right) \nonumber 
   \\
   &+\left(\frac{\xi ^3}{8}+\frac{3 \xi
   ^2}{4}+\frac{73 \xi }{24}+\frac{275}{36}\right) C_A^2+\left(\frac{1}{3}-\frac{\xi
   }{2}\right) C_F N_f-\frac{3 \xi  C_F^2}{4}+\frac{2 N_f^2}{9}\bigg]  \nonumber 
   \\
   & +\frac{C_F}{\epsilon}
   \bigg[C_A \left(\Big(\frac{143}{12}-4 \zeta _3\Big) C_F+\Big(\frac{17 \xi
   }{24}+\frac{287}{54}\Big) N_f\right)-\frac{C_F N_f}{2}-\frac{C_F^2}{2} -\frac{5
   N_f^2}{27}\nonumber 
   \\
   & +C_A^2 \bigg\{-\frac{5 \xi ^3}{48}+\xi ^2
   \left(-\frac{\zeta _3}{8}-\frac{13}{32}\right)+\xi  \left(-\frac{\zeta
   _3}{4}-\frac{263}{96}\right)+\frac{23 \zeta
   _3}{8}-\frac{9155}{432}\bigg\}\bigg] \,,
   \\
 Z_q^{(4)}\big|_{N_f^3} =& \frac{C_F}{9 \epsilon ^3}-\frac{5 C_F}{54 \epsilon ^2}-\frac{35 C_F}{324 \epsilon }\,, 
 \\ 
    Z_q^{(4)}\big|_{N_f^2}  =&\frac{\xi  C_F C_A}{18 \epsilon
   ^4}  + \frac{C_F}{\epsilon ^3} \bigg[\left(-\frac{\xi }{6}-\frac{23}{12}\right) C_A+\left(\frac{1}{6}-\frac{2 \xi }{9}\right) C_F\bigg]\nonumber
   \\
   & +\frac{C_F}{\epsilon ^2} \bigg[\left(\frac{2 \xi
   }{27}+4\right) C_A+\left(\frac{5 \xi }{27}+\frac{3}{8}\right) C_F\bigg] \nonumber
   \\
   & +\frac{C_F}{{\epsilon }}\bigg[C_A \left(\left(\frac{269}{972}-\frac{\zeta
   _3}{3}\right) \xi -2 \zeta _3-\frac{293}{144}\right)+\left(2 \zeta _3-\frac{19}{9}\right) C_F\bigg].
   \end{align}
   We note that at four loops, only the $N_f^3$ and $N_f^2$ terms contribute to the calculation presented in this paper.

\bibliographystyle{JHEP}
\bibliography{nf2sp}

\end{document}